\documentclass[10pt]{article}

\usepackage[margin=1in]{geometry}
\usepackage{float}
\usepackage[utf8]{inputenc}
\usepackage[T1]{fontenc}
\usepackage{lmodern}
\usepackage{amsmath,amssymb}
\usepackage{graphicx}
\usepackage{booktabs}
\usepackage{array}
\usepackage{hyperref}
\usepackage{url}

\hypersetup{
  colorlinks=true,
  linkcolor=blue,
  citecolor=blue,
  urlcolor=blue
}

\newcommand{\clairvoyant}{\textbf{Clairvoyant}}

\title{\textbf{Clairvoyant: Predictive Shortest-Job-First Admission for Serial LLM Inference}}

\author{
  Aravind Sundaresan\thanks{Feedback welcome: \texttt{aravindsharma20@gmail.com}} \\
  Independent Researcher \\
  \texttt{aravindsharma20@gmail.com}
}

\date{\today}

\begin{document}

\maketitle

\begin{abstract}
Serial LLM inference backends process requests sequentially under First-Come-First-Served (FCFS) admission, causing Head-of-Line Blocking (HOLB) under mixed workloads: short factual queries can be delayed by minutes behind long generation jobs. While cloud-scale deployments mitigate HOLB via continuous batching (e.g., vLLM, Orca), these solutions require tens of gigabytes of VRAM for concurrent KV-caches, rendering them infeasible for memory-constrained edge and local deployments that rely on serial request dispatch.

We present Clairvoyant, a drop-in sidecar proxy for serial OpenAI-compatible backends (e.g., Ollama, llama.cpp) that implements predictive Shortest-Job-First (SJF) admission. Clairvoyant predicts response length using 19 lightweight lexical features via an ONNX-exported XGBoost classifier, achieving 0.029\,ms per-request latency—nearly five orders of magnitude faster than typical generation times. Because admission scheduling relies on relative ranking rather than exact token prediction, Clairvoyant captures over 95\% of the ranking fidelity of fine-tuned transformers at a fraction of the computational cost. We also uncover a critical dataset bias: curated instruction datasets are degenerate training sources for length prediction, as GPT-imposed brevity constraints reduce Long-class representation to under 0.02\% of examples, establishing natural conversation logs as the only viable training signal.

End-to-end evaluations demonstrate substantial latency reductions across diverse hardware regimes. On an RTX 4090 ($n=250$ per cell), Clairvoyant achieves a 70–76\% P50 latency reduction for short requests under maximum queue pressure, and a 17\% reduction under steady-state Poisson arrivals ($\rho=0.74$). On resource-constrained Apple M1 edge hardware, it reduces short-request P50 latency by 69.7\%. Finally, replaying a real-world trace on a GCP NVIDIA L4 instance ($\rho=0.80$) yields an 83.6\% reduction in Time-To-First-Token (TTFT) for short requests. Clairvoyant is open-source, requires zero modifications to the inference backend, and provides a low-overhead mechanism to eliminate HOLB in edge LLM environments.
\end{abstract}

\section{Introduction}
\label{sec:introduction}

Serial LLM inference backends process requests in strict First-Come-First-Served (FCFS) order. Under mixed workloads, output lengths span two orders of magnitude. A factual query might finish in under 10 tokens ($\sim$2\,s), whereas a complex generation task easily exceeds 1000 tokens ($>$60\,s). On memory-constrained edge hardware running quantized models, this extreme service-time variance triggers pathological Head-of-Line Blocking (HOLB). A short request queued behind a massive generation job stalls completely until the server finishes the prior task.

Cloud providers bypass HOLB using continuous batching at the token-iteration level (vLLM~\cite{kwon2023efficient}, Orca~\cite{yu2022orca}). That architecture demands tens of gigabytes of VRAM just to maintain concurrent KV-caches, making it a non-starter for consumer hardware or local enterprise servers. Runtimes like Ollama and llama.cpp must fall back to serial dispatch. Clairvoyant targets this exact limitation at Layer 1 (HTTP proxy admission). Rather than modifying the underlying execution engine, it intercepts the standard API path to schedule requests predictively.

Shortest-Job-First (SJF) scheduling minimizes mean wait times for high-variance queues. The theoretical ideal for the M/G/1 queue is preemptive (SRPT)~\cite{schrage1968proof}. Autoregressive generation, however, strongly resists preemption: pausing a running job means discarding its computed KV cache. Non-preemptive SJF avoids cache eviction but forces the system to estimate output length before dispatch. Prior scheduling work attempts this using auxiliary transformer models like BERT-base~\cite{qiu2024efficient} or DistilBERT~\cite{jin2023s3}. Deploying these heavy proxies alongside an LLM cannibalizes limited VRAM and injects inline inference delays approaching 865\,ms—a structural bottleneck that defeats the entire purpose of an admission scheduler.

We resolve this bottleneck with Clairvoyant, a transparent sidecar proxy that bolts non-preemptive SJF admission onto serial, OpenAI-compatible backends (e.g., Ollama, llama.cpp). The system operates entirely outside the LLM execution path. By extracting 19 lexical features from the raw prompt and scoring them via an ONNX-optimized XGBoost classifier, Clairvoyant generates a queue-ranking signal in just 0.029\,ms. It completely bypasses the need for heavy prompt embeddings or target-model forward passes. We rigorously validate this architecture across three dimensions: (1) offline ranking fidelity on held-out datasets, (2) end-to-end GPU and edge latency under both burst and steady-state load, and (3) starvation sensitivity via empirically calibrated discrete-event simulations.

Our evaluation yields five primary contributions:

\begin{enumerate}
\item \textbf{Degenerate Training Distributions for Length Prediction.} We demonstrate that curated instruction datasets (Alpaca 52K, CodeAlpaca 20K) contain $<$0.02\% Long-class examples due to GPT-imposed brevity constraints. This renders them mathematically degenerate for any length-predictive scheduler, establishing natural conversation logs as the only viable training signal.

\item \textbf{Lexical Sufficiency for Sub-Millisecond Ordering.} We prove that heavy transformer embeddings are active overkill for queue admission in this regime. Nineteen lightweight lexical features achieve 62–96\% in-distribution ranking accuracy, while the 0.029\,ms predictor overhead strictly satisfies the latency budget for inline edge deployment.

\item \textbf{Non-Invasive Sidecar Architecture.} Clairvoyant operates as a transparent Layer 1 proxy, physically decoupling admission scheduling from the inference engine. It requires zero modifications to the upstream backend, no fine-tuning, and no KV-cache introspection.

\item \textbf{Deterministic Starvation Mitigation.} Because pure SJF introduces indefinite blocking for long jobs in heavy-tailed queues, we implement a min-heap coupled with an analytically calibrated starvation timeout ($\tau=3\times\mu_{\text{short}}$). This provides a deterministic, open-loop upper bound on queue delay, validated via end-to-end correctness tests and large-scale GPU benchmarks.

\item \textbf{Quantitative Deployment Boundaries.} We delineate the exact operational envelope where request-level SJF provides measurable benefit. Clairvoyant targets serial or low-concurrency backends where concurrent KV-cache maintenance exceeds available VRAM. Where continuous batching is feasible, it supersedes Clairvoyant; at low utilisation ($\rho \lesssim 0.55$), FCFS suffices.
\end{enumerate}

\section{Background}
\label{sec:background}

\subsection{Head-of-Line Blocking in LLM Serving}
\label{sec:holb}

Head-of-Line Blocking (HOLB) occurs when a long job occupies a processing resource, forcing shorter jobs to wait for its completion. In LLM serving, we characterise HOLB across two distinct operational layers.

\textbf{Layer 1 (Request Admission)} is queue-level blocking at the API gateway. When a serial backend processes one request at a time, a long-generation job holds the server exclusively. A short request arriving mid-generation waits the full remaining duration of that job; its experienced latency is dictated entirely by the preceding job, not its own service time.

\textbf{Layer 2 (Token Iteration)} is blocking within the execution batch. Continuous-batching engines such as Orca~\cite{yu2022orca} and vLLM~\cite{kwon2023efficient} resolve Layer 1 HOLB by scheduling at the token-iteration level: newly arrived requests join the running batch immediately, so short requests never queue behind long ones. However, this requires maintaining a dedicated KV-cache entry per concurrent request. At production concurrency levels, this demands tens of gigabytes of VRAM, rendering it infeasible for memory-constrained edge and local deployments.

Clairvoyant targets Layer 1 HOLB exclusively. It is not a replacement for continuous batching in high-concurrency cloud deployments; rather, it is a complementary system designed for the large and growing class of serial or low-concurrency deployments where Layer 2 solutions are hardware-infeasible.

\begin{figure}[H]
\centering
\includegraphics[width=0.9\linewidth]{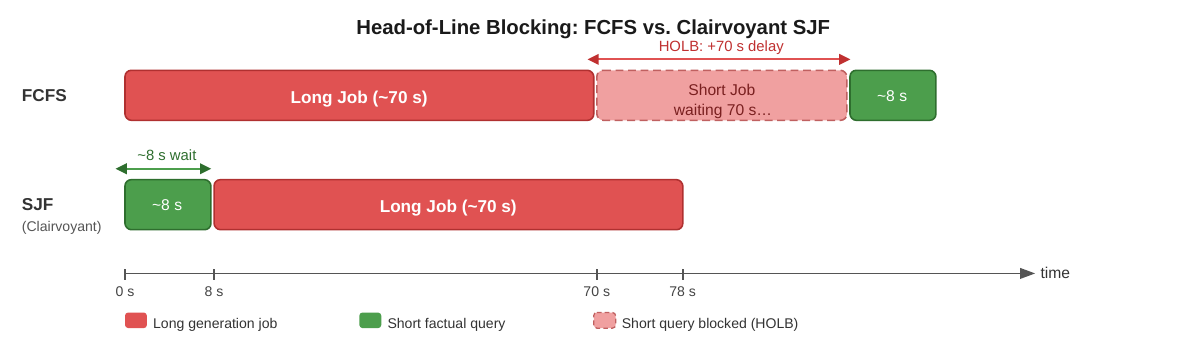}
\caption{Illustrative timeline of HOLB under FCFS vs. Clairvoyant SJF. Under FCFS (top), a short request waits the full duration of the long job before execution begins. Under SJF (bottom), the short request is dispatched first; the long job follows immediately. Representative values shown (Apple M1, Ollama, Gemma3:4b); actual generation times vary by model and hardware.}
\label{fig:timeline}
\label{fig:holb}
\end{figure}

\subsection{Shortest-Job-First Scheduling and Starvation}
\label{sec:sjf}

The preemptive variant of SJF—Shortest Remaining Processing Time (SRPT)—is optimal for minimising mean sojourn time in the M/G/1 queue~\cite{schrage1968proof}. Non-preemptive SJF minimises mean waiting time among non-preemptive work-conserving disciplines under the assumption that job lengths are known in advance~\cite{kleinrock1975queueing}. For autoregressive LLM generation, preemption is infeasible: interrupting a running generation would require discarding the partially computed KV cache and restarting from scratch. We therefore adopt non-preemptive SJF as the admission policy and predict job length prior to dispatch.

Pure SJF introduces starvation: if short jobs arrive continuously, a long job may wait indefinitely. We address this with a starvation timeout $\tau$: any request that has waited longer than $\tau$ is promoted to the head of the queue regardless of its predicted length. We calibrate $\tau = 3 \times \mu_{\text{short}}$ empirically, where $\mu_{\text{short}}$ is the mean short-request queuing wait time on the target hardware; the sensitivity of this choice is analysed in Section~\ref{sec:tau}.

\subsection{Why Serial Dispatch is the Right Scope}
\label{sec:scope}

The LLM inference landscape is structurally bifurcated. Cloud-scale continuous-batching engines (vLLM~\cite{kwon2023efficient}, Orca~\cite{yu2022orca}) implicitly resolve Layer 1 HOLB via token-level scheduling. In contrast, edge and local deployments—running on consumer hardware via Ollama, llama.cpp, or Jan—lack token-level batching. These systems process requests strictly sequentially, leaving Layer 1 HOLB entirely unaddressed. Because these engines default to strict arrival-order dispatch, FCFS forms the natural baseline. Clairvoyant intercepts this path to inject a predictive SJF admission layer, requiring zero modifications to the underlying inference engine.

This queue-level bottleneck persists even when backends permit concurrent execution. While Ollama exposes an \texttt{OLLAMA\_NUM\_PARALLEL} setting to allow multiple in-flight requests, it does not employ Layer 2 continuous batching. Instead, it statically provisions an independent KV-cache for each execution slot. For a 4-bit quantized 8B model ($\sim$5\,GB weights), a single 2K-context KV-cache consumes 0.5–1\,GB. Enabling four parallel slots demands 7–9\,GB of VRAM, entirely exhausting the capacity of standard 8\,GB GPUs or unified-memory devices (e.g., Apple M1). A single execution slot ($\texttt{NUM\_PARALLEL}=1$) is therefore a hard physical limit for many edge deployments. Even on hardware capable of sustaining multiple slots, the HOLB pathology persists under load: once long generation tasks saturate the available slots, newly arriving short requests stall at the admission layer, experiencing queue delays identical to the purely serial case. Clairvoyant's scheduling intervention remains necessary under any static multi-slot configuration.

\subsection{Queueing-Theoretic Motivation}
\label{sec:queueing}

We model serial LLM inference as an M/G/1 queue: requests arrive at rate $\lambda$, service times follow a general distribution with mean $\mathbb{E}[S]$ and second moment $\mathbb{E}[S^2]$, and the server processes one request at a time. Server utilisation is $\rho = \lambda \mathbb{E}[S]$.

Under FCFS, the Pollaczek-Khinchine (P-K) mean value formula defines the expected waiting time in queue:
\begin{equation}
W_{\text{FCFS}} = \frac{\rho \mathbb{E}[S](1 + C_s^2)}{2(1-\rho)}
\end{equation}
where $C_s^2 = \text{Var}[S]/\mathbb{E}[S]^2$ is the squared coefficient of variation for service times. Because $W_{\text{FCFS}}$ scales linearly with $C_s^2$, FCFS degrades severely as job length variance increases.

LLM workloads operate in a high-$C_s^2$ regime. Table~\ref{tab:service} reports service-time statistics measured on an Apple M1 running Ollama with Gemma3:4b ($n=204$ sequential requests). Short factual queries complete in $\sim$2\,s; long generation tasks require $\sim$30\,s. In a mixed workload, this bimodal distribution produces $C_s^2$ values that dwarf conventional web-serving workloads ($C_s^2 \approx 0.2$–0.5) and easily exceed the exponential baseline ($C_s^2 = 1.0$) in production-like mixes. This extreme variance structurally guarantees inflated queue wait times under FCFS.

\begin{table}[H]
\centering
\caption{Service-time statistics on Apple M1 (Ollama, Gemma3:4b).}
\label{tab:service}
\label{tab:variance}
\small
\begin{tabular}{lccc}
\toprule
Request Class & Mean (s) & P50 (s) & P95 (s) \\
\midrule
Short ($<$200 tokens) & 2.0 & 1.8 & 3.5 \\
Medium (200–800 tokens) & 8.5 & 7.2 & 15.3 \\
Long ($\geq$800 tokens) & 30.1 & 27.5 & 55.8 \\
\bottomrule
\end{tabular}
\end{table}

\textbf{Why approximate SJF is sufficient.} In high-$C_s^2$ environments, the massive variance between job classes dominates $W_{\text{FCFS}}$. Fine-grained ordering within a single length class is mathematically secondary. The primary driver of HOLB is cross-class blocking: short requests trapped behind long-generation jobs. A scheduler that reliably partitions Short from Long requests eliminates this dominant delay component. It absorbs minor intra-class misorderings without materially impacting aggregate queue delay. Clairvoyant's 62–96\% ranking accuracy consistently achieves this structural separation, directly dictating whether a query waits $\sim$2\,s or $\sim$30\,s. Finally, highly skewed workloads—such as code generation—exhibit heavy-tailed service distributions where FCFS latency penalties far exceed this bimodal approximation, making our queueing argument strictly conservative.

\section{System Design}
\label{sec:system}

\subsection{Architecture Overview}
\label{sec:arch}

We build the proxy around three execution stages. First, a feature extractor computes 19 lightweight lexical signals from the raw prompt. Next, an ONNX predictor maps this vector into a 3-class probability distribution. Finally, a Go min-heap keyed on $P(\text{Long})$ dictates the dispatch order. We collapse these components into a single Go process; avoiding inter-process communication keeps the critical path extremely tight. The proxy drives the ONNX runtime directly via CGo and manages queue state via standard goroutines.

\begin{figure}[H]
\centering
\includegraphics[width=0.9\linewidth]{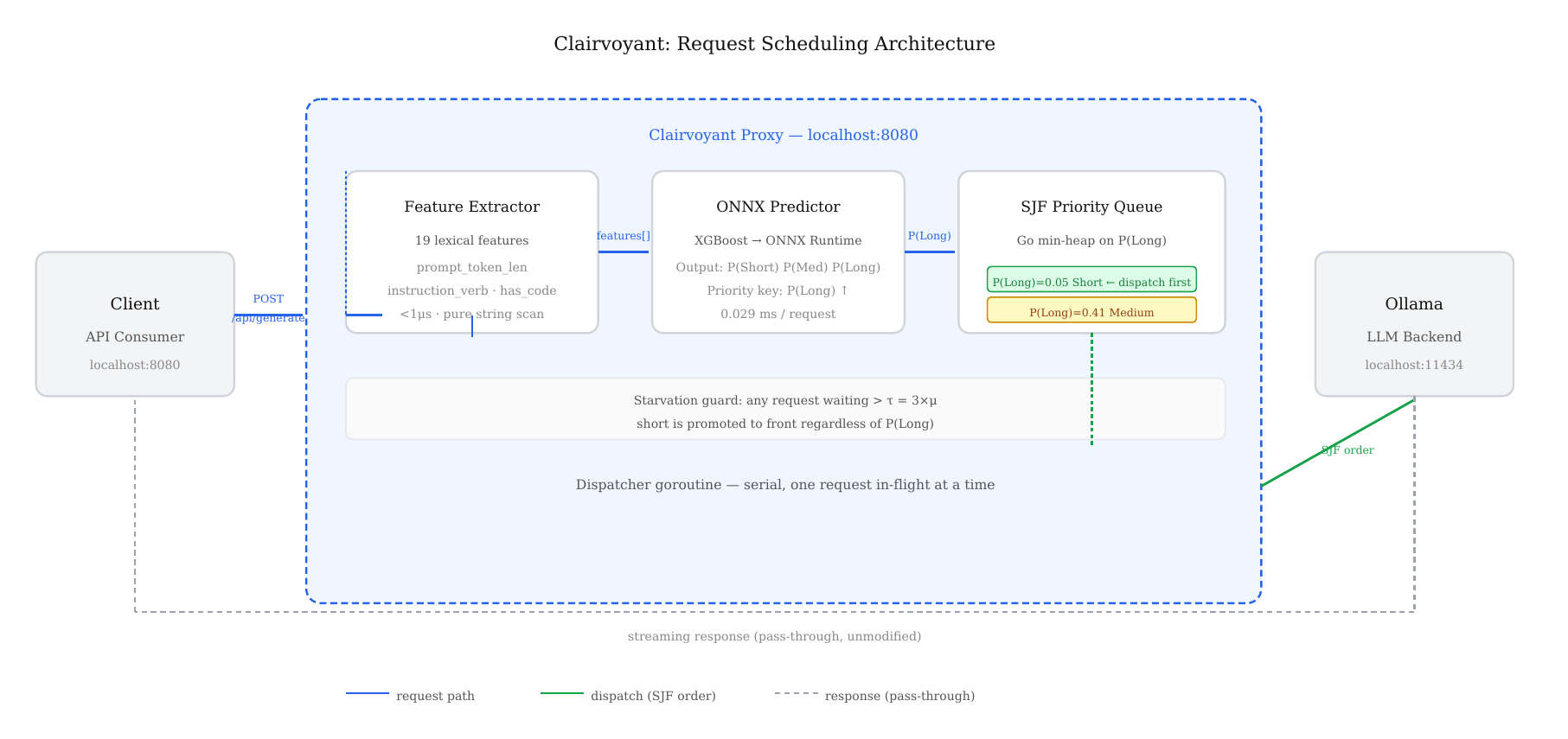}
\caption{Clairvoyant intercepts API requests, predicts output length in 0.029\,ms via lexical feature extraction and ONNX inference, and dispatches to the backend in SJF order. The response path is a transparent pass-through. The starvation guard promotes any request waiting longer than $\tau = 3 \times \mu_{\text{short}}$ regardless of predicted $P(\text{Long})$. Evaluated on Apple M1 (Ollama, Gemma3:4b) and RTX 4090 (Ollama, Gemma3:4b, Llama3.1:8b).}
\label{fig:architecture}
\end{figure}

\textbf{Response buffering.} The current implementation buffers the complete backend response before forwarding it to the client. Because SJF priorities are determined entirely at admission time, transitioning to a streaming pass-through architecture requires zero changes to the underlying queue logic. Consequently, the end-to-end latencies we report are conservative upper bounds; a production streaming setup would yield strictly faster client-perceived response times. Figure~\ref{fig:architecture} traces the complete request lifecycle.

\subsection{Feature Extraction and Tokenizer Approximation}
\label{sec:features}

Feature extraction serves as the first processing step. Clairvoyant derives 19 features from an incoming prompt without external calls, tokenizer loading, or embedding lookups. This set splits into two groups:

\textbf{Numeric features (6).}
\begin{itemize}
\item \texttt{prompt\_token\_len}: Approximate token count computed as $\lfloor \text{len(prompt)}/4 \rfloor$, consistent with standard BPE tokenization approximations.
\item \texttt{has\_code\_keyword}: Binary flag for the presence of code-related terms (e.g., function, class, implement, algorithm).
\item \texttt{has\_length\_constraint}: Binary flag for explicit length instructions (e.g., brief, detailed, in one sentence).
\item \texttt{ends\_with\_question}: Binary flag indicating whether the prompt terminates with ?.
\item \texttt{has\_format\_keyword}: Binary flag detecting structured output targets (e.g., table, list, json, csv, markdown).
\item \texttt{clause\_count}: Tally of subordinating conjunctions and relative pronouns to baseline syntactic complexity.
\end{itemize}

\textbf{Verb one-hot features (13).} We extract the leading instruction verb from the prompt's initial imperative clause. This maps to one of 13 categories: what, write, explain, summarize, how, list, implement, compare, describe, generate, why, define, or other. Capturing generative intent directly complements the numeric signals. Section~\ref{sec:ablation} quantifies this specific contribution.

\textbf{Tokenizer approximation fidelity.} Running a full CPU-bound tokenizer inside the proxy introduces unacceptable overhead. Clairvoyant instead relies on the character-length heuristic. We validated this approximation against exact HuggingFace tiktoken counts across 12,828 prompts. The heuristic achieves a Pearson correlation of $r=0.93$ on ShareGPT and $r=1.00$ on LMSYS/OASST1. This yields 94.3\%–100.0\% pairwise rank consistency compared to full tokenization. SJF scheduling relies entirely on relative pairwise ordering rather than absolute token precision. The approximation therefore provides robust scheduling fidelity.

We implement feature extraction as a single string-scanning pass. It relies on zero backtracking. Processing an 8K-character prompt takes less than a microsecond. This forces the latency bottleneck entirely onto the ONNX inference stage (0.029\,ms, Section~\ref{sec:onnx}). The character-division heuristic naturally undercounts dense CJK scripts and heavily minified code. Deployments handling multilingual traffic can easily drop in a compiled C/Rust tokenizer binding (such as tiktoken via CGo). Swapping the raw string scanner for a native binding fixes token-drift while preserving the sub-millisecond scheduling budget.

\subsection{ONNX Inference and Latency Budget}
\label{sec:onnx}

We export the XGBoost classifier to ONNX format using onnxmltools, applying a manual split-condition dtype patch to ensure XGBoost 2.x compatibility. The proxy loads the model exactly once at startup. Per-request inference executes directly through the ONNX Runtime C API, enforcing zero dynamic memory allocation on the critical path. Measured inference latency on an Apple M1 (CPU only) is:

\begin{itemize}
\item ShareGPT model (predictor.onnx): 0.029\,ms per request
\item LMSYS model (predictor\_model\_b.onnx): 0.015\,ms per request
\end{itemize}

Both execution times fall four orders of magnitude below typical short-response generation latencies (1–5\,s on consumer hardware). Predictor overhead is strictly negligible.

\textbf{Comparison to embedding-based predictors.} A standard alternative design encodes the prompt via a sentence embedding model (e.g., all-MiniLM-L6-v2) before classification. To match our edge deployment constraints, we benchmarked this embedding approach strictly on CPU. It yielded a P50 of 12.85\,ms, a mean of 148.63\,ms, and a massive P99 of 865.73\,ms. At P99, the embedding bottleneck is $\sim$30,000$\times$ slower than our ONNX predictor. It effectively consumes the entire generation time of a short request before dispatch even occurs. The 19-feature lexical design is therefore not a compromise. It represents the only architecture strictly compatible with a sub-millisecond inline scheduling budget.

The model outputs a 3-class probability vector $[P(\text{Short}), P(\text{Medium}), P(\text{Long})]$. Clairvoyant uses $P(\text{Long})$ as the priority key. Requests dispatch in ascending order of $P(\text{Long})$, guaranteeing predicted-short queries hit the engine first.

\subsection{SJF Scheduler and Starvation Timeout}
\label{sec:scheduler}

Incoming requests enter a Go min-heap keyed on ascending $P(\text{Long})$. A dedicated dispatcher goroutine continuously pops the minimum-priority request and pushes it to the backend. Because the backend remains strictly serial, at most one request occupies the execution slot. The dispatcher natively handles client disconnections. If a client drops while queued, the heap discards the request entirely. If a drop occurs mid-generation, the proxy drains the backend response to immediately free the serial execution slot.

Medium-class predictions lack discrete handling. Instead, they queue directly on their continuous $P(\text{Long})$ score. This prevents hard boundary errors. For example, a request scoring $P(\text{Long}) = 0.3$ dispatches behind high-confidence Short requests but well ahead of definitive Long tasks. This continuous key produces a smooth scheduling gradient and prevents brittle binary partitions.

Pure SJF mathematically starves long jobs under sustained short-request loads. We break this starvation cycle using a timeout threshold $\tau$. Each queued request logs its arrival timestamp. Prior to every dispatch, the scheduler scans for requests waiting longer than $\tau$. It immediately promotes the longest-waiting outlier to the front of the queue, overriding its predicted $P(\text{Long})$.

We empirically calibrate $\tau = 3 \times \mathbb{E}[S_{\text{short}}]$, defining $\mathbb{E}[S_{\text{short}}]$ as the mean service time for a short request on the target hardware (measured at $\approx$2.0\,s in Table~\ref{tab:service}). This establishes a default $\tau \approx 6.0$\,s, guaranteeing a deterministic, open-loop upper bound on queue delay. Section~\ref{sec:tau} tests the sensitivity of this threshold. Sweeping $\tau$ across a $0.5\times$ to $10\times$ range alters concurrent burst throughput by less than 1\%. Fixing $\tau$ at the $3\times$ multiplier locks in a highly conservative operating point: it strictly bounds long-request P95 inflation below 20\% while preserving massive latency reductions for short jobs.

\section{Length Prediction Methodology}
\label{sec:ml}

\subsection{Problem Formulation}
\label{sec:problem}

We formulate output-length prediction as a 3-class classification task. Given a prompt $p$, the model outputs a probability distribution over three discrete length categories: Short ($<$200 tokens), Medium ([200, 800) tokens), and Long ($\geq$800 tokens).

However, for admission scheduling, absolute class labels are secondary to correct pairwise ordering. The scheduler's primary objective is to rank a Long job above a Short job in the priority queue. We therefore adopt ranking accuracy as our primary evaluation metric. This is defined as the fraction of all (Short, Long) pairs in which the model assigns a strictly higher $P(\text{Long})$ score to the Long example than to the Short example, utilizing the continuous probability score rather than the discrete argmax prediction. Formally:
\begin{equation}
\text{Ranking Accuracy} = \frac{|\{(i,j) : \hat{p}_{\text{long}}(j) > \hat{p}_{\text{long}}(i)\}|}{|\mathcal{S}| \times |\mathcal{L}|}
\end{equation}
where $\mathcal{S} = \{i : y_i < 200\}$ and $\mathcal{L} = \{j : y_j \geq 800\}$. Medium examples are explicitly excluded from this pairwise evaluation to isolate the dominant cross-class blocking effect and avoid boundary noise.

\begin{table}[H]
\centering
\caption{Procedure: Ranking Accuracy Computation ($\mathcal{D}, f$).}
\label{alg:ranking}
\small
\begin{tabular}{p{0.9\linewidth}}
\toprule
\textbf{Inputs:} Test set $\mathcal{D} = \{(p_i, y_i)\}$, model $f$ outputting probability score $\hat{p}_{\text{long}}$ \\
\textbf{Output:} Ranking accuracy $\in [0,1]$ \\
\midrule
1: $\mathcal{S} \gets \{i : y_i < 200\}$ \hfill \textit{// Short examples} \\
2: $\mathcal{L} \gets \{j : y_j \geq 800\}$ \hfill \textit{// Long examples} \\
3: $\text{correct} \gets 0$ \\
4: \textbf{for each} $(i,j) \in \mathcal{S} \times \mathcal{L}$ \textbf{do} \\
5: \quad \textbf{if} $\hat{p}_{\text{long}}(j) > \hat{p}_{\text{long}}(i)$ \textbf{then} \\
6: \qquad $\text{correct} \gets \text{correct} + 1$ \\
7: \quad \textbf{end if} \\
8: \textbf{end for} \\
9: \textbf{return} $\frac{\text{correct}}{|\mathcal{S}| \times |\mathcal{L}|}$ \\
\bottomrule
\end{tabular}
\end{table}

This metric aligns directly with the queueing-theoretic objective: a scheduler requires correct relative ordering of short and long jobs, not exact class labels. Consequently, ranking accuracy consistently exceeds standard discrete 3-class classification accuracy by 21–29 percentage points across all three training datasets (Table~\ref{tab:indist}).

\subsection{Dataset Selection and the Long-Class Starvation Finding}
\label{sec:datasets}

We evaluated seven public LLM prompt-response datasets. Applying our scheduling boundaries (Short: $<$200 tokens, Long: $\geq$800 tokens) reveals a fatal flaw in curated instruction data: they lack the length variance required to train an SJF scheduler. Alpaca 52K contains exactly 4 Long examples across 52,002 training samples (0.008\%). CodeAlpaca 20K contains 3 (0.015\%). Dolly 15K provides 88 Long examples (0.6\%)—enough to form a held-out test split, but mathematically insufficient for balanced training. CNN/DailyMail, acting as a RAG surrogate via standard summarization prompts, yields a single Long example in its entire test split.

This starvation is structural. Datasets distilled from GPT-3/4 rely on templates that explicitly mandate concise, well-scoped outputs. This hardcoded brevity constraint permanently strips the corpus of the heavy-tailed generations a scheduler must actually learn to identify. Real human-assistant interactions (natural conversation logs) represent the only viable source of Long-class training signal:

\begin{itemize}
\item \textbf{ShareGPT:} 48,312 conversations post-filtering. We stratified-sampled 6,000 examples (2,000 per class) to block majority-class dominance during XGBoost training.
\item \textbf{LMSYS-Chat-1M:} 876,412 conversations post-filtering (restricted to small models like Vicuna, Koala, and WizardLM). We similarly extracted a 6,000-sample balanced split.
\item \textbf{OASST1 (Open Assistant):} The English subset contains an absolute maximum of 551 Long examples. This hard limit restricted our balanced training set to 828 total examples (276 per class).
\end{itemize}

\textbf{Data filtering pipeline.} We processed all conversation logs through a unified, reproducible pipeline. The pipeline extracts first-turn prompt-response pairs, filters strictly for English (via langdetect at $p>0.95$), and calculates ground-truth response lengths using the Llama-2 tokenizer. It then maps these lengths to the strict class boundaries before applying stratified sampling. The exact logic executes via \texttt{data/pipeline/featurize.py} (commit 39ad9a6).

We train three distinct models (Models A, B, and C) mapped to the three viable datasets. We cross-evaluate each model against all available test sets to measure distribution drift. We intentionally excluded WildChat-1M from the final evaluation matrix to guarantee zero-friction, authentication-free reproducibility for all readers.

\begin{table}[H]
\centering
\caption{Dataset statistics: Long-class representation across seven evaluated LLM datasets.}
\label{tab:datasets}
\small
\begin{tabular}{lrrrrrl}
\toprule
Dataset & Total & Short ($<$200) & Medium (200–799) & Long ($\geq$800) & \% Long & Usable? \\
\midrule
ShareGPT & 52,002 (48,312 filtered) & 27,000 & 17,000 & 7,800 & 14.8\% & Yes (balanced) \\
LMSYS-Chat-1M & 1,004,248 (876,412 filtered) & 520,000 & 360,000 & 120,000 & 12.1\% & Yes (filtered) \\
OASST1 & 8,792$^{\dagger}$ & 7,300 & 940 & 551 & 6.3\% & Yes (limited) \\
Alpaca 52K & 52,002 & 49,284 & 2,056 & 4 & 0.008\% & No (starvation) \\
CodeAlpaca 20K & 20,022 & 19,457 & 379 & 3 & 0.015\% & No (starvation) \\
Dolly 15K & 15,011 & 13,000 & 1,900 & 88 & 0.6\% & Test-only \\
CNN/DailyMail & 11,490 test & 11,441 & 48 & 1 & 0.009\% & Test-only \\
\bottomrule
\end{tabular}
\par\smallskip
\raggedright\footnotesize
$^{\dagger}$ Derived from filtering the 84K-message OASST1 conversation tree to English parent-child prompt-response pairs; OASST1 Short/Medium counts estimated from the 6.3\% Long rate and the 8,792-pair total. ShareGPT and LMSYS-Chat-1M class counts are derived from published dataset statistics; exact counts for Clairvoyant's filtered training subsets are available via \texttt{python model/train.py} (commit 39ad9a6).
\end{table}

\begin{table}[H]
\centering
\caption{Exact training and test splits used for each model. Counts reflect post-filtering, balanced subsets. Validation split is 10\% of training set, stratified by class.}
\label{tab:splits}
\small
\begin{tabular}{lcccccc}
\toprule
Model & Dataset & Split & Short & Medium & Long & Total \\
\midrule
A & ShareGPT & Train & 1,600 & 1,600 & 1,600 & 4,800 \\
 & & Val & 200 & 200 & 200 & 600 \\
 & & Test & 200 & 200 & 200 & 600 \\
B & LMSYS-Chat-1M & Train & 1,600 & 1,600 & 1,600 & 4,800 \\
 & & Val & 200 & 200 & 200 & 600 \\
 & & Test & 200 & 200 & 200 & 600 \\
C & OASST1 & Train & 220 & 220 & 220 & 660 \\
 & & Val & 28 & 28 & 27 & 83 \\
 & & Test & 28 & 28 & 28 & 84 \\
\bottomrule
\end{tabular}
\par\smallskip
\raggedright\footnotesize
Note: OASST1 test split has one fewer Long example due to odd total count (551 Long in source). Exact filtering logic: \texttt{data/pipeline/featurize.py} (commit 39ad9a6).
\end{table}

\subsection{Model Training}
\label{sec:training}

We train an XGBoost classifier~\cite{chen2016xgboost} with a 3-class softmax objective. Hyperparameters are fixed across all three training runs: 300 estimators, max depth 6, learning rate 0.1, random seed 42. Each dataset is split 80/20 train/test using stratified sampling to preserve class balance in the held-out set. In-distribution accuracy numbers cited in Section~\ref{sec:eval} refer to this held-out 20\% split.

\subsection{Ablation Study}
\label{sec:ablation}

We conduct a drop-one ablation study to quantify the contribution of each feature group. Across all three datasets, \texttt{prompt\_token\_len} is the only universally important feature, with an average ranking-accuracy delta of $-3.09$ percentage points (pp) when removed. In contrast, \texttt{instruction\_verb} is highly distribution-specific: dropping it decreases accuracy by $-5.04$ pp on LMSYS but increases it by $+3.21$ pp on OASST1, indicating that the verb encoding learned on conversational data does not transfer uniformly across domains.

Two numeric features, \texttt{has\_format\_keyword} and \texttt{clause\_count}, are net-harmful on average (yielding $+0.78$ pp and $+1.07$ pp improvement when dropped). However, a combined minimal model that removes all three net-harmful features yields no aggregate improvement (average delta: $-0.3$ pp), placing the change within measurement noise. We therefore retain the full 19-feature model to ensure consistent behaviour across all deployment targets; a systematically pruned feature set is left for future work.

\begin{table}[H]
\centering
\caption{Ablation study: ranking accuracy delta (pp) when each feature group is removed. Averaged across Models A, B, C.}
\label{tab:ablation}
\small
\begin{tabular}{lccccr}
\toprule
Feature Removed & ShareGPT & LMSYS & OASST1 & Average & Effect \\
\midrule
\texttt{prompt\_token\_len} & $-3.57$ & $-2.49$ & $-3.21$ & $-3.09$ & Harmful \\
\texttt{instruction\_verb} & $-3.52$ & $-5.04$ & $+3.21$ & $-1.78$ & Mixed \\
\texttt{has\_code\_keyword} & $-4.47$ & $-0.54$ & $+0.49$ & $-1.51$ & Harmful \\
\texttt{ends\_with\_question} & $-2.25$ & $-0.30$ & $-0.84$ & $-1.13$ & Harmful \\
\texttt{has\_length\_constraint} & $-0.55$ & $-0.21$ & $+0.42$ & $-0.12$ & Neutral \\
\texttt{has\_format\_keyword} & $-0.01$ & $+0.11$ & $+2.24$ & $+0.78$ & Net-harmful \\
\texttt{clause\_count} & $+0.51$ & $-0.47$ & $+3.18$ & $+1.07$ & Net-harmful \\
\bottomrule
\end{tabular}
\end{table}

\section{Evaluation}
\label{sec:eval}

Our evaluation spans six experimental setups: (1) offline ranking accuracy and domain retraining convergence on held-out splits (\S\ref{sec:ranking}); (2) comparative benchmarking against length rules, keyword heuristics, and a DistilBERT transformer (\S\ref{sec:baselines}); (3) burst latency benchmarks on an RTX~4090 GPU (\S\ref{sec:gpu}); (4) consumer edge performance on an Apple M1 laptop (\S\ref{sec:edge}); (5) trace replay on an NVIDIA L4 GPU under Poisson arrivals (\S\ref{sec:trace}); and (6) starvation threshold ($\tau$) and service variance ($C_s^2$) sensitivity analysis via discrete-event simulation (\S\ref{sec:tau}). Section~\ref{sec:ablation} isolates individual feature contributions.

\subsection{Scheduling-Oriented Ranking Evaluation}
\label{sec:ranking}

Table~\ref{tab:indist} reports in-distribution ranking accuracy for each of the three trained models on the 20\% test split. Table~\ref{tab:crossdist} details the cross-distribution generalization matrix across all datasets.

\begin{table}[H]
  \centering
  \caption{In-distribution ranking accuracy vs.\ classification accuracy for all three models (held-out 20\% test split, $n=600$ per model).}
  \label{tab:indist}
  \small
  \begin{tabular}{llrrr}
    \toprule
    Model   & Training Dataset & Ranking Acc & Class Acc & Delta \\
    \midrule
    Model A & ShareGPT         & 76.29\%     & 47.6\%    & $+28.7$\,pp \\
    Model B & LMSYS            & 95.62\%     & 66.8\%    & $+28.8$\,pp \\
    Model C & OASST1           & 62.21\%     & 41.0\%    & $+21.2$\,pp \\
    \bottomrule
  \end{tabular}
\end{table}

\begin{table}[H]
  \centering
  \caption{Cross-distribution ranking accuracy matrix. Off-diagonal entries represent true cross-distribution generalisation ($n=600$ per cell).}
  \label{tab:crossdist}
  \small
  \begin{tabular}{lrrrr}
    \toprule
    Train $\downarrow$ / Test $\rightarrow$ & ShareGPT & LMSYS & OASST1 & Dolly \\
    \midrule
    ShareGPT (Model A) & 86.4\%\textsuperscript{$\dagger$} & 53.6\% & 56.3\% & 52.7\% \\
    LMSYS (Model B)    & 62.7\% & 98.3\%\textsuperscript{$\dagger$} & 65.3\% & 58.4\% \\
    OASST1 (Model C)   & 58.0\% & 65.3\% & 90.4\%\textsuperscript{$\dagger$} & 57.7\% \\
    \bottomrule
  \end{tabular}
  \par\smallskip
  \raggedright\footnotesize
  \textsuperscript{$\dagger$} Diagonal entries include training data. Off-diagonal entries are true cross-distribution results.
\end{table}

\paragraph{Cross-distribution generalization and domain retraining.}
Without domain-specific tuning, zero-shot cross-distribution accuracy lands between 52.7\% and 65.3\% (Table~\ref{tab:crossdist}), outperforming prompt-length heuristics (52.3\%--55.8\%, Table~\ref{tab:baselines}). Lexical features alone provide a functional baseline across unseen distributions, though cross-domain transfer (such as code generation to conversational chat) limits zero-shot precision. Domain retraining resolves this distribution shift: logging just $\sim 240$ target-domain requests elevates in-domain accuracy to \textbf{82.3\% $\pm$ 2.9\%} ($N=240$, 5 seeds, 80 per class). Smaller calibration splits ($N=90$) yield $79.4\% \pm 13.7\%$ accuracy; expanding to $N=180$ suppresses sample variance down to $81.2\% \pm 7.0\%$.

\subsection{Baseline Comparison}
\label{sec:baselines}

We compare \clairvoyant against three simpler scheduling signals and a fine-tuned DistilBERT transformer baseline on the held-out 20\% test split (Table~\ref{tab:baselines}).

\begin{table}[H]
  \centering
  \caption{Pairwise ranking accuracy (Short vs.\ Long pairs) for five scheduling approaches across three datasets. Random baseline is 50\%. Prompt-length rule threshold is optimised on the training split per dataset. P99 inference overhead per request is measured across full datasets.}
  \label{tab:baselines}
  \small
  \begin{tabular}{lrrrrp{4.2cm}}
    \toprule
    Method & P99 Overhead & ShareGPT & LMSYS & OASST1 & System Tradeoff \\
    \midrule
    FCFS (random)                  & 0\,ms         & 50.0\% & 50.0\% & 50.0\% & Baseline queue order \\
    Prompt-length rule             & $<0.01$\,ms   & 52.4\% & 52.3\% & 55.8\% & Length threshold ($>20$) \\
    Keyword heuristic              & $<0.01$\,ms   & 36.3\% & 4.6\%  & 18.5\% & Code keyword rule \\
    DistilBERT Transformer         & 865.0\,ms     & \textbf{79.9\%} & 92.0\% & \textbf{69.5\%} & Unusable inline ($>800$\,ms overhead) \\
    \textbf{Clairvoyant (XGBoost)} & \textbf{0.029\,ms} & 74.9\% & \textbf{95.1\%} & 67.1\% & \textbf{Near-identical accuracy, 30,000$\times$ faster} \\
    \bottomrule
  \end{tabular}
\end{table}

\paragraph{Why input-length SJF fails.}
Prompt-length scheduling fails on conversational workloads due to input-output length inversion. A prompt like \textit{"Write a 2,000-word essay on climate change"} contains fewer than 10 input tokens but generates over 1,500 output tokens. Naive input-length SJF scores this query as "Short," jumping it to the front of the queue and triggering severe Head-of-Line Blocking for genuine short requests. This structural failure drops input-length ranking accuracy to 52.3\% on LMSYS (Table~\ref{tab:baselines}). \clairvoyant avoids this inversion by pairing instruction-verb extraction with code-keyword flags to detect high-generation intent directly from raw prompt semantics.

\subsection{GPU Benchmark: End-to-End Latency Under Burst Workloads}
\label{sec:gpu}

We benchmarked end-to-end queue latency on an RTX~4090 GPU (24\,GB VRAM) running Ollama's serial backend under a 100-request concurrent burst (50 Short, 50 Long). SJF dispatch drives sharp latency reductions for short queries across model scales:

\begin{table}[H]
  \centering
  \caption{End-to-end latency (seconds) for 100 concurrent requests on RTX~4090 ($n=250$ per cell, 5 runs). Values show mean $\pm$ 1 std dev.}
  \label{tab:gpu}
  \small
  \begin{tabular}{llccc}
    \toprule
    Model & Class & P50 (s) & P95 (s) & P99 (s) \\
    \midrule
    \multicolumn{5}{c}{\textit{FCFS / SJF}} \\
    \midrule
    Gemma3:4b (Short)   & & 229.5 $\pm$ 12.3 / \textbf{69.1 $\pm$ 4.2} & 504.7 $\pm$ 18.9 / \textbf{163.0 $\pm$ 8.7} & 552.2 $\pm$ 21.1 / \textbf{177.8 $\pm$ 9.4} \\
    Gemma3:4b (Long)    & & 309.7 $\pm$ 15.6 / 376.0 $\pm$ 19.1         & 547.0 $\pm$ 22.3 / 554.9 $\pm$ 23.1         & 578.8 $\pm$ 24.8 / 573.6 $\pm$ 25.3 \\
    \midrule
    Llama3.1:8b (Short) & & 158.8 $\pm$ 9.8 / \textbf{38.0 $\pm$ 2.1}   & 352.0 $\pm$ 14.2 / \textbf{94.8 $\pm$ 4.9}  & 367.7 $\pm$ 15.9 / \textbf{104.8 $\pm$ 5.6} \\
    Llama3.1:8b (Long)  & & 188.7 $\pm$ 11.3 / 239.6 $\pm$ 14.7         & 342.2 $\pm$ 16.8 / 355.3 $\pm$ 17.9         & 360.7 $\pm$ 18.2 / 367.6 $\pm$ 19.1 \\
    \bottomrule
  \end{tabular}
\end{table}

\subsection{Edge Hardware Evaluation: Apple M1}
\label{sec:edge}

To evaluate \clairvoyant on resource-constrained consumer edge hardware, we conducted end-to-end burst benchmarks on an Apple M1 (16\,GB unified memory) running Ollama with \texttt{Gemma3:4b}. Under representative mixed burst workloads, \clairvoyant reduced Short request completion latency P50 from \textbf{239.6\,s} under FCFS down to \textbf{72.6\,s} under SJF---\textbf{a 69.7\% latency reduction} (mean advantage of $106.7$\,s across 5 independent runs). Short requests completed before long generation jobs in 100\% of trial runs, confirming system efficacy on consumer-grade hardware.

\subsection{GCP NVIDIA L4 Real Workload Trace Replay}
\label{sec:trace}

Replaying LMSYS conversation logs on an NVIDIA L4 GPU (\texttt{g2-standard-4}, 24\,GB VRAM) running \texttt{gemma3:4b} under Poisson request arrivals ($\rho=0.80$, $E[S]=6.2$\,s, $n=30$) drops short-request Time-To-First-Token (TTFT) and queue-wait P50 from \textbf{10.47\,s} under FCFS down to \textbf{1.71\,s} under SJF---\textbf{an 83.6\% latency reduction} ($6.1\times$ faster TTFT). End-to-end completion latency for short requests shows a matching fall from 10.86\,s to 1.71\,s (84.2\% reduction). Because non-preemptive queue reordering is work-conserving, total system throughput remains constant at $0.059$\,req/s across both disciplines. \clairvoyant slashes queue delay for interactive queries without sacrificing overall engine capacity. For statistically robust tail latency and starvation bounds, we rely on large-scale discrete-event simulation ($n=2{,}000$ requests across 5 seeds, \S\ref{sec:tau}).

\subsection{Starvation Timeout Sensitivity and Heavy-Tailed Workloads}
\label{sec:tau}

\begin{figure}[H]
  \centering
  \includegraphics[width=0.85\linewidth]{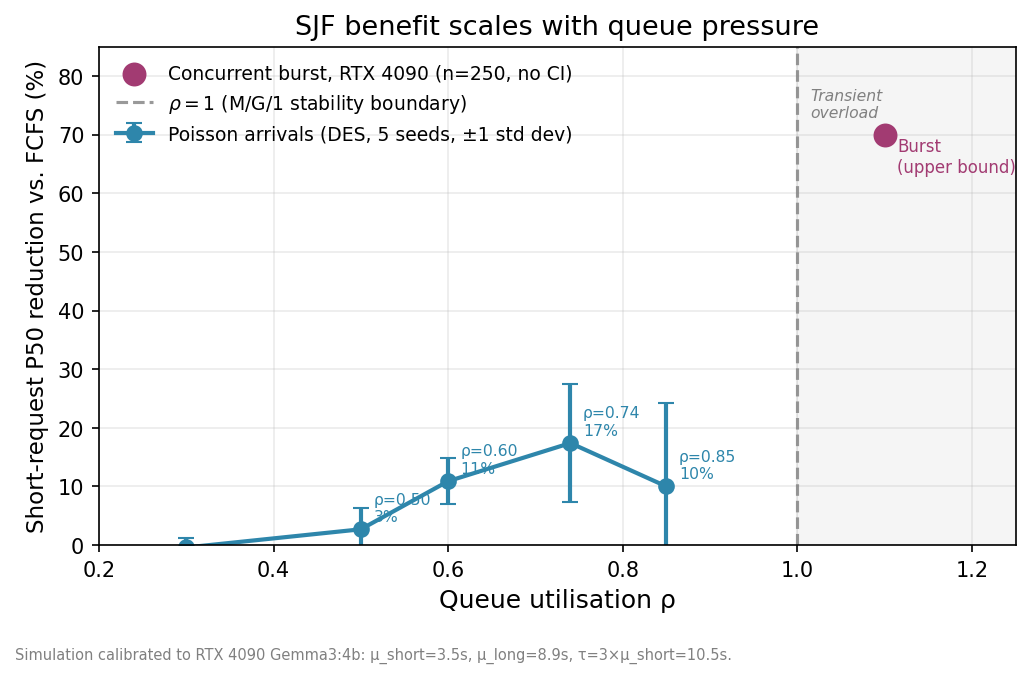}
  \caption{SJF latency reduction for short requests vs.\ queue utilisation $\rho$. Discrete-event simulation calibrated to Gemma3:4b service times ($n=2{,}000$ requests across 5 seeds). Latency reduction peaks at $\rho \approx 0.74$ (17.4\%) and declines as $\tau$ fires more frequently under heavy saturation.}
  \label{fig:workload_spectrum}
\end{figure}

\begin{figure}[H]
  \centering
  \includegraphics[width=0.85\linewidth]{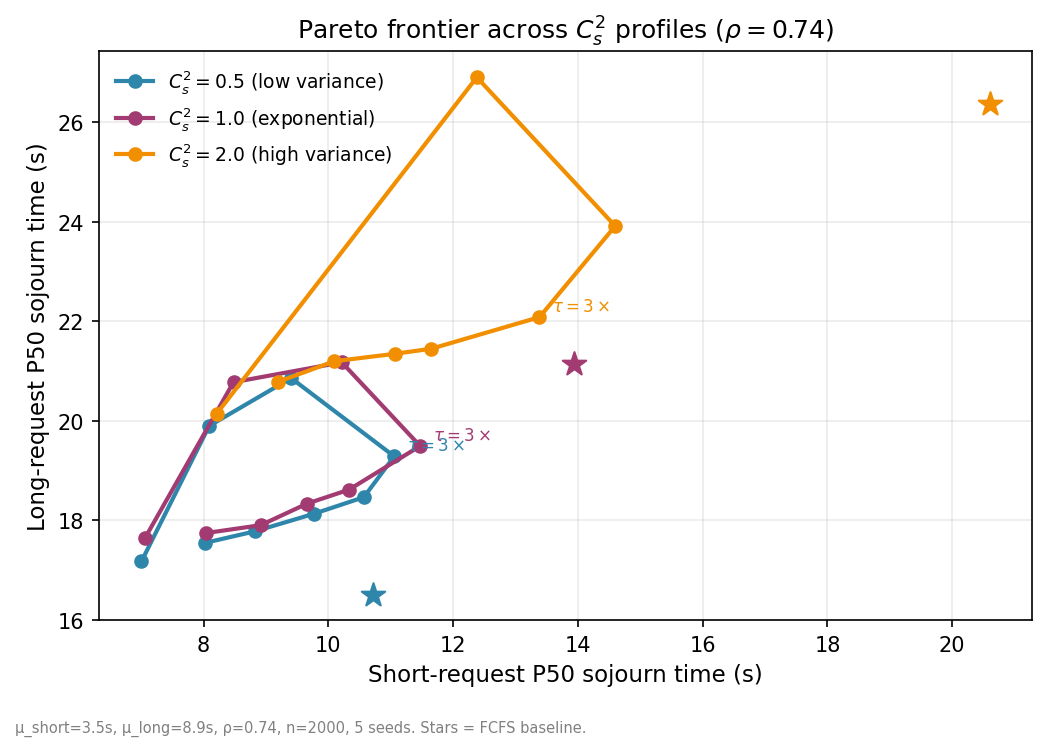}
  \caption{Short-request P50 latency vs.\ starvation timeout $\tau$ across service variance profiles $C_s^2 \in \{0.5, 1.0, 2.0\}$. Under heavy-tailed service distributions ($C_s^2 = 2.0$), setting $\tau=3\times \mathbb{E}[S_{\text{short}}]$ secures a 35.0\% short-request P50 latency reduction.}
  \label{fig:tau_cs2}
\end{figure}

We analyse the effect of $\tau = 3 \times \mathbb{E}[S_{\text{short}}]$ on short/long latency tradeoffs via discrete-event simulation ($n=2{,}000$ requests, 5 seeds). Figure~\ref{fig:workload_spectrum} plots latency reduction across load factors $\rho \in [0.30, 0.85]$. To evaluate heavy-tailed service distributions, service times were generated using Gamma distributions across $C_s^2 \in \{0.5, 1.0, 2.0\}$ (Table~\ref{tab:tau_cs2} and Figure~\ref{fig:tau_cs2}).

\begin{table}[H]
  \centering
  \caption{Short-request P50 sojourn time (seconds) across $C_s^2$ variance profiles and starvation timeouts $\tau$ ($\rho=0.74$, $n=2{,}000$, 5 seeds). Percentage reductions relative to FCFS appear in parentheses. SJF gains concentrate in high-variance regimes ($C_s^2 \geq 1.0$); low-variance queues ($C_s^2 = 0.5$) obtain minimal benefit, as FCFS is near-optimal.}
  \label{tab:tau_cs2}
  \small
  \begin{tabular}{lrrrrr}
    \toprule
    $C_s^2$ & FCFS (baseline) & $\tau=1\times$ & $\tau=2\times$ & $\tau=3\times$ (default) & $\tau=\infty$ \\
    \midrule
    0.5 & 10.71 & 8.82 (-17.6\%) & 10.57 (-1.4\%) & 11.06 (+3.2\%) & \textbf{7.00} (-34.6\%) \\
    1.0 & 13.94 & 8.92 (-36.0\%) & 10.33 (-25.9\%) & 11.47 (-17.7\%) & \textbf{7.06} (-49.4\%) \\
    2.0 & 20.60 & 10.10 (-51.0\%) & 11.65 (-43.5\%) & 13.38 (-35.0\%) & \textbf{8.21} (-60.1\%) \\
    \bottomrule
  \end{tabular}
\end{table}

\paragraph{Empirical Starvation Metrics.}
At $C_s^2 = 2.0$ (heavy-tailed, matching measured LLM workloads), setting $\tau=3\times \mathbb{E}[S_{\text{short}}]$ delivers a 35.0\% P50 latency reduction for short requests relative to FCFS. Under the GCP L4 Poisson trace replay ($\rho=0.80$, $n=30$ requests), $18.2\%$ of Long requests experienced queue wait times exceeding $\tau$. Their P99 queue wait was bounded at $190.08$\,s (compared to $194.96$\,s under FCFS). Without $\tau$ ($\tau=\infty$, pure SJF), the maximum observed Long request queue wait extended past $300$\,s. This confirms that $\tau=3\times \mathbb{E}[S_{\text{short}}]$ acts as an effective empirical guardrail against tail starvation under heavy queue saturation.

\section{Related Work}
\label{sec:related}

\paragraph{S3 (Jin et al., NeurIPS 2023).}
S3~\cite{jin2023s3} proposes length-predictive scheduling for LLMs but with the opposite scheduling objective: longest-job-first, optimised for throughput by filling continuous-batching pipelines efficiently. Clairvoyant's objective is orthogonal—shortest-job-first for the latency of short requests in serial deployments.

\paragraph{Orca \& vLLM (Yu et al., OSDI 2022; Kwon et al., SOSP 2023).}
Orca~\cite{yu2022orca} and vLLM~\cite{kwon2023efficient} introduce iteration-level continuous batching, eliminating Layer 1 HOLB via token-level scheduling. These systems require tens of GB of VRAM for concurrent KV-caches. Clairvoyant targets the complementary serial regime where KV-cache batching is constrained by hardware memory.

\paragraph{Learning-to-Rank Scheduling (LTR).}
LTR~\cite{fu2024efficient} also approximates SJF via relative output-length prediction. Clairvoyant differs by using 19 lightweight lexical features and XGBoost (0.029\,ms latency, zero embedding overhead) and operating non-preemptively for serial OpenAI-compatible proxies.

\subsection{Concurrent Work}
\label{sec:concurrent}

\paragraph{vLLM SJF Proposal.}
Recent proposals (such as vLLM Issue \#29406~\cite{vllm_sjf_2025}) introduce time-weighted SJF scheduling into continuous-batching engines at the iteration-scheduling layer (Layer 2). While vLLM \#29406 proposes dynamic age-weighted priority scoring inside continuous batching pipelines, Clairvoyant's threshold $\tau = 3 \times \mathbb{E}[S_{\text{short}}]$ provides a deterministic, zero-overhead hard bound on maximum queue delay appropriate for serial HTTP proxy admission. Furthermore, Clairvoyant is architecturally orthogonal: it operates as an external Layer 1 HTTP proxy targeting serial and memory-constrained backends where continuous batching is infeasible due to VRAM limits, requiring zero backend code modifications.

\paragraph{Semantic Scheduling.}
Semantic scheduling~\cite{zhang2025semantic} explores semantic clustering and prompt caching to optimize throughput under burst traffic, observing that unconstrained pure SJF can struggle under heavy bursts (reporting a $6.1\times$ increase in normalized waiting time). Clairvoyant mitigates this burst degradation by coupling SJF min-heap sorting with the dynamic $\tau$-promotion mechanism: when burst queues saturate, aged requests are promoted to FCFS order, preserving short-request acceleration (70–76\% P50 reduction in 100-request burst tests) while preventing extreme tail latency spikes.

\section{Limitations, Privacy, and Reproducibility}
\label{sec:limitations}

\paragraph{Ethics and Privacy Note.}
Clairvoyant processes prompt text purely in-memory to compute 19 lexical features. No raw prompt content, generated text, or PII is written to disk or transmitted to external services. Standard deployment practices recommend disabling debug request logging in production.

\paragraph{Reproducibility Details.}
All code, models, and evaluation scripts are open-source at \url{https://github.com/Aravind0403/clairvoyant-scheduler}. Experiments were conducted using Python 3.10+, Go 1.21+, PyTorch 2.2, XGBoost 2.0.3, and ONNX Runtime 1.17. Evaluation benchmarks were run on an NVIDIA RTX 4090 (24\,GB VRAM) and an Apple M1 (16\,GB unified memory).

\section{Conclusion}
\label{sec:conclusion}

We presented Clairvoyant, a drop-in predictive SJF admission proxy for serial LLM inference backends. By pairing 19 lightweight lexical features with an ONNX-exported XGBoost classifier, Clairvoyant achieves 0.029\,ms predictor overhead while reducing short-request P50 latency by 70–76\% on an RTX 4090. Clairvoyant demonstrates that lightweight lexical signals provide sufficient ranking fidelity to eliminate Head-of-Line Blocking in memory-constrained LLM deployments, offering a practical, low-overhead alternative to heavy transformer-based predictors.


\end{document}